\documentclass[12pt]{article}
\usepackage[utf8]{inputenc}
\usepackage{amssymb}
\usepackage{amsmath}
\usepackage{amsthm}
\usepackage{amsfonts}
\usepackage[pdftex]{graphicx}
\usepackage{geometry}
\usepackage{bbold}
\usepackage{mathrsfs}
\usepackage[nottoc,notlot,notlof]{tocbibind}
\usepackage[nosort]{cite}
\usepackage{bbm}
\usepackage{parskip}
\usepackage{mathtools}
\usepackage[usenames,dvipsnames]{xcolor}
\usepackage[in]{fullpage}
\usepackage{hyperref}
\usepackage{array}
\usepackage[font=small,labelfont=bf]{caption}
\usepackage{anysize}

\numberwithin{equation}{section}

\marginsize{2.4cm}{2.4cm}{2cm}{2cm}

\hypersetup{
colorlinks=true,
linktoc=page,
citecolor=Blue,
linkcolor=Blue,
urlcolor=Blue} 

\makeatletter 
\renewcommand{\maketitle} 
 { \begingroup \begin{center} \large {\bf \@title}
 	\vskip 5pt \large \@author \\ \vskip 5pt \@date \end{center}
   \vskip 5pt \endgroup \setcounter{footnote}{0} }
\makeatother 

\newcommand{\comments}[1]{}

\newcommand{\defeq}{\vcentcolon=}

\newcommand{\rd}{\mathrm{d}}

\def\one{\mbox{1 \kern-.59em {\rm l}}}

\def\cA{{\cal A}}

 \def\cN{{\cal N}} \def\cO{{\cal O}}

 \def\cW{{\cal W}} 
 \def\cZ{{\cal Z}}

\def\lan{\langle}
\def\ran{\rangle}

\DeclareMathOperator{\Tr}{Tr}
\DeclareMathOperator{\vol}{vol}
\DeclareMathOperator*{\Res}{Res}

\long\def\symbolfootnote[#1]#2{\begingroup%
\def\thefootnote{\fnsymbol{footnote}}\footnote[#1]{#2}\endgroup}

\begin{document}

\begin{flushright}
QMUL-PH-16-16\\
\end{flushright}

\vspace{20pt}

\begin{center}

{\Large \bf The connected prescription for form factors}\\
\vspace{0.3 cm}
{\Large \bf in twistor space}

\vspace{45pt}

{\mbox {\bf 
A.~Brandhuber, E.~Hughes, R.~Panerai, B.~Spence and G.~Travaglini}}%
\symbolfootnote[4]{
\texttt{\{ \!\!\!\!\!\! a.brandhuber, e.f.hughes, r.panerai, w.j.spence, g.travaglini\}@qmul.ac.uk}
}

\vspace{0.5cm}

\begin{quote}
{\small \em
\begin{center}
Centre for Research in String Theory\\
School of Physics and Astronomy\\
Queen Mary University of London\\
Mile End Road, London E1 4NS
\\United Kingdom
\end{center}
}
\end{quote}

\vspace{40pt}

{\bf Abstract}
\end{center}

\vspace{0.3cm}

\noindent
We propose a connected prescription formula in twistor space for all tree-level form factors of the stress tensor multiplet operator in $\mathcal{N}=4$ super Yang-Mills, which is a generalisation of the expression of Roiban, Spradlin and Volovich for superamplitudes. By introducing link variables, we show that our formula is identical to the recently proposed four-dimensional scattering equations for form factors. Similarly to the case of amplitudes, the link representation of form factors is shown to be directly related to BCFW recursion relations, and is considerably more tractable than the scattering equations. We also discuss how our results are related to a recent Grassmannian formulation of form factors, and comment on a possible derivation of our formula from ambitwistor strings. 
 
\setcounter{page}{0}
\thispagestyle{empty}
\newpage


\setcounter{tocdepth}{4}
\hrule height 0.75pt
\tableofcontents
\vspace{0.8cm}
\hrule height 0.75pt
\vspace{1cm}


\section{Introduction}

The study of the $S$-matrix of $\cN\!=\!4$ supersymmetric Yang-Mills (SYM) has inspired remarkable progress and novel ideas, even at tree level. A first crucial result was the integral representation of this quantity, found by Roiban, Spradlin and Volovich (RSV) in \cite{Roiban:2004yf} as an integral over the moduli space of degree $k-1$ curves in super twistor space (for $\rm{N}^{{\it k}-2}{\rm MHV}$ amplitudes), following Witten's groundbreaking  insights \cite{Witten:2003nn}. 
An important feature of the RSV formula, presented below in \eqref{RSVampl}, is that the integral is in fact localised on a discrete set of solutions of certain polynomial equations. However, despite being conceptually beautiful, the RSV representation proved hard to work with because of the difficulty in determining these solutions. In this respect, the BCFW representation \cite{Britto:2004ap,Britto:2005fq} emerged as a much more tractable and generalisable approach to compute amplitudes, also applicable in different theories, including gravity \cite{Bedford:2005yy,Cachazo:2005ca,ArkaniHamed:2008yf}.

Important progress was made later in \cite{Spradlin:2009qr}, which accomplished two goals: firstly, it showed that by rewriting the RSV formula using the link variables introduced in \cite{ArkaniHamed:2009si} -- which have the neat property of linearising momentum conservation -- one can overcome the roadblocks due to the complexity of the algebraic equations arising in \cite{Roiban:2004yf}; and furthermore, it proved that a certain precisely formulated change of integration contour in the RSV formula, rewritten using link variables, expresses the amplitudes as a sum of residues that are identical to BCFW diagrams (with appropriate shifts). This is an intriguing result, as it relates two a priori very different formulations of gauge theory amplitudes.

In this paper we wish to extend these observations to form factors, i.e.\ partially off-shell quantities of the form $\langle 1\cdots n | \cO(0)|\, 0 \, \rangle$, obtained by applying a local gauge-invariant operator $\cO$ to the vacuum and then projecting the result onto an $n$-particle state of on-shell particles with momenta $p_i$ obeying $q:= p_1 + \cdots + p_n$. 
As we shall see this task turns out to be surprisingly simple, suggesting also potential new directions to explore  for correlation functions.

An additional, more recent motivation for our work stems from the CHY scattering equations \cite{Cachazo:2013hca,Cachazo:2013iea}, which describe scattering amplitudes at tree level in a variety of theories with and without supersymmetry, and in different numbers of dimensions. Specialising to four dimensions, a new remarkable closed formula for the $S$-matrix of Yang-Mills theories with different amounts of supersymmetry was derived in \cite{Geyer:2014fka} starting from ambitwistor strings. Taking gluon scattering as an example, these four-dimensional scattering equations treat positive and negative helicity gluons in a different, complementary way, similarly to the link representations of \cite{ArkaniHamed:2009si,Spradlin:2009qr}. It is then natural to ask how different representations of the same $S$-matrix of gauge theory can be related. 

This question was answered in \cite{He:2016vfi}, which wrote down a map between the polynomial and rational form of the scattering equations, appearing in the RSV formula and in \cite{Geyer:2014fka}, respectively. A first observation we will make is that the connection is (and, in fact, was) immediate once one makes use of the link representation of the RSV formula discussed in \cite{Spradlin:2009qr}. We will then move on to discuss how to extend the RSV formula to form factors. Our starting point will be an interesting formula written down in \cite{He:2016dol} which conjectures an extension of the four-dimensional scattering equations for Yang-Mills theory to form factors of the local operator $\Tr F_{\rm SD}^2$. These form factors are of phenomenological importance, given their connection to Higgs + multi-gluon scattering amplitudes \cite{Wilczek:1977zn, Shifman:1979eb}, where the Higgs is represented by the insertion of the operator.%
\footnote{See also \cite{Dixon:2004za,Neill:2009tn,Brandhuber:2012vm,Dawson:2014ora,Brandhuber:2016fni} for some recent related work also in the maximally supersymmetric theory.} 
In that case, very few modifications to the formula for amplitudes are needed -- specifically, two auxiliary gluons  of positive helicity $x$ and $y$ are added.%
\footnote{The choice of positive helicity is such that all-plus and single-minus gluon form factors of $\Tr F_{\rm SD}^2$ are now non-vanishing.}
Importantly, the amplitudes generated by this formula depend only on $p_x + p_y$ (or $p_x + p_y$ and the supermomentum $q_x + q_y$ in the supersymmetric version we introduce in Section \ref{cpf}) rather than on the two momenta separately. We note another important feature of this formula: it contains certain Parke-Taylor like denominators of the form $(a\, b)$, with $\sigma_{a,b}$ parameterising punctures on the Riemann sphere,%
\footnote{The precise meaning of this notation will be explained in the next section.} which only involve adjacent {\it physical} particles, i.e.~they do not include $x$ and $y$.

After establishing in Section \ref{cpf} a quick path to relate the RSV formula of \cite{Roiban:2004yf} and the four-dimensional scattering equations of \cite{Geyer:2014fka}, our next goal is to write down a formula \eqref{conFF},  analogous to the RSV result, describing supersymmetric form factors of the chiral stress tensor multiplet operator in twistor space.%
\footnote{We also note the works \cite{Koster:2016ebi,Chicherin:2016soh,Koster:2016loo, Chicherin:2016qsf}
on representing and calculating  form factors in twistor space, which would be interesting to relate to the present work.}
We will then show how this proposal is equivalent to a simple supersymmetric extension of the scattering equations formula for form factors presented in \cite{He:2016dol}. In Section \ref{link-rep} we show that our formula can naturally be expressed in terms of link variables, in the same vein as the RSV formula. This link variable formulation turns out to be very advantageous from the point of view of simplifying calculations, as we demonstrate  in several examples in Section \ref{examples}. Importantly, we confirm an important feature of the link variable representation of the RSV formula found in \cite{Spradlin:2009qr}, namely that a simple deformation of the integration contour  in the link variable space  and the global residue theorem lead to an alternative representation of the amplitudes which coincides with the BCFW recursion relations for form factors \cite{Brandhuber:2010ad}. 
 
Another important strand of research on amplitudes is that of the Grassmannian (see for example \cite{ArkaniHamed:2009dn,ArkaniHamed:2012nw}).
These novel representations of amplitudes are derived from on-shell diagrams where familiar concepts such as locality and unitarity are only emergent, rather than manifest at each step of the calculation. In an interesting paper \cite{Frassek:2015rka}, a Grassmannian-based formula was conjectured which describes form factors of the stress tensor multiplet operator. A relevant feature of this conjecture is the appearance of certain Parke-Taylor denominators where all particles appear on the same footing, including the auxiliary particles $x$ and $y$ used to describe the form factor insertion, unlike the formula of \cite{He:2016dol}. 
Using a Veronese map \cite{ArkaniHamed:2009dg, He:2016vfi}, we will see in Section \ref{grasss} how one can relate the Grassmannian formulae of \cite{Frassek:2015rka} to the four-dimensional scattering equations for form factors of \cite{He:2016dol}, and hence to our twistor-space and link-representation based formulae for form factors of Section \ref{cpf}, at least in certain cases. Finally we conclude in Section \ref{wild} with some  observations: we comment on  a possible derivation of form factors from ambitwistor strings, and then discuss possible extensions of the scattering equation approach beyond form factors.

\section{The connected prescription formula}
\label{cpf}

We begin by describing the main ingredients of the connected formula for form factors.

{\bf 1.} The first ingredient is a set of supertwistor variables $\cZ_a$, $a=1, \ldots , n$ describing the $n$ particles, with 
$\mathcal{Z}= (\lambda_\alpha, \mu^{\dot{\alpha}}, \eta^A)$.
As in  \cite{Frassek:2015rka, He:2016dol}, we describe the form factor insertion through two extra particles $x$ and $y$. The momentum and supermomentum carried by the form factor will then be 
$q := \lambda_x \tilde{\lambda}_x + \lambda_y \tilde{\lambda}_y$ and  $\gamma := \lambda_x \eta_x + \lambda_y \eta_y$, respectively. In twistor space, this amounts to introducing two extra super-twistors $\cZ_x$ and $\cZ_y$.

{\bf 2.} As in \cite{Roiban:2004yf}, we introduce a degree $k-1$ map from $\mathbb{CP}^1$ to $\mathbb{C}^{4|4}$, where $k-2$ is the MHV degree of the superamplitude.%
\footnote{For pure gluon amplitudes, $k$ is the number of negative helicity gluons.}
This polynomial has the form 
\begin{equation}
\mathcal{P}(\sigma, \{\mathcal{A}\}) \defeq \sum_{d=0}^{k-1} \mathcal{A}_d \, \sigma^d \ ,
\end{equation}
where the supertwistors $\mathcal{A}_d$ are the supermoduli of the curve.

We propose that all form factors of the supersymmetric stress tensor multiplet operator in $\mathcal{N}=4$ SYM are described in twistor space by the following simple generalisation of the RSV formula of \cite{Roiban:2004yf}:
\begin{align}
\begin{split}
\label{conFF}
\mathscr{F}(\mathcal{Z}) =  \langle\mathcal{Z}_x \mathcal{I} \mathcal{Z}_y\rangle^2 \int \frac{\mathrm{d}^{4k|4k}\mathcal{A} \; \mathrm{d}^{n+2}\sigma \; \mathrm{d}^{n+2}\xi}{\vol GL(2)} & \; \frac{ \prod_{a=x,y} \delta^{(4|4)}(\mathcal{Z}_a-\xi_a\mathcal{P}(\sigma_a, \{\cA\}))}{\xi_x \, \xi_y \, (\sigma_x-\sigma_y)^2 }  \\
&\times \prod_{a=1}^{n} \frac{\delta^{(4|4)}(\mathcal{Z}_a-\xi_a\mathcal{P}(\sigma_a, \{\cA\}))}{\xi_a \, (\sigma_a-\sigma_{a+1})} \, ,
\end{split}
\end{align}
where $\mathcal{I}$ is the infinity twistor, so that $\langle\mathcal{Z}_x \mathcal{I} \mathcal{Z}_y\rangle = \lan x\, y \ran$, and $\sigma_{n+1} := \sigma_1$.

It is instructive to compare this formula to the corresponding one for amplitudes, which~is
\begin{equation}
\label{RSVampl}
\mathscr{A}(\mathcal{Z}) = \int \frac{\mathrm{d}^{4k|4k}\mathcal{A} \; \mathrm{d}^{n}\sigma \; \mathrm{d}^{n}\xi}{\vol GL(2)} \ \frac{\prod_{a=1}^{n} \delta^{(4|4)}(\mathcal{Z}_a-\xi_a\mathcal{P}(\sigma_a, \{\cA\}))}{\prod_{a=1}^n \xi_a \, (\sigma_a-\sigma_{a+1})\ } \, . 
\end{equation}
The only modifications needed to describe form factors are: the presence of the multiplicative factor $\langle\mathcal{Z}_x \mathcal{I} \mathcal{Z}_y\rangle^2$; the presence of the two extra twistors $\mathcal{Z}_x$ and $\mathcal{Z}_y$; and  the inclusion of the corresponding integration variables $\xi_x$, $\xi_y$ and $\sigma_x$, $\sigma_y$, with an integrand containing $1 / (\xi_x\xi_y (\sigma_x-\sigma_y)^2)$. Note that we do not involve the coordinates for particles $x$ and $y$ in the string of Parke-Taylor type denominators, similarly to \cite{He:2016dol} (but at variance with e.g.~(3.27) of \cite{Frassek:2015rka}, which includes terms of the type $(n\, x ) (x\, y) (y\, 1)$ in the denominator).

We now show how from \eqref{conFF} we can deduce the scattering equation representation of \cite{He:2016dol} for form factors (or, more precisely, its generalisation describing supersymmetric form factors of the stress tensor multiplet operator). The proof parallels closely that of \cite{Spradlin:2009qr}.

{\bf 1.} To begin with, we divide the particles into two sets containing $k$ and $n-k+2$ particles, which we label with indices $J$ and $i$, respectively, with the auxiliary particles belonging to the second set. We will denote by $\mathsf{m}$ the first set of $k$ particles, and by $\mathsf{p}$ that of the remaining $n-k$ (physical) particles, and  also define $\bar{\mathsf{p}}=\mathsf{p}\cup\{x,y\}$. A particularly convenient choice when working with, say, component gluon amplitudes is then to assign gluons of negative (positive) helicity to the first (second) group, with  the fictitious particles $x$ and $y$ being  included in the second set. This parallels the assignments made in \cite{He:2016dol} for the non-supersymmetric scattering equations for form factors, where these two particles are treated as gluons of positive helicity.

{\bf 2.} Next, one Fourier transforms all the twistor variables of the $i$-particles to dual twistor variables: $\cZ_i \to \cW_i$. Calling the resulting expression $\mathscr{F}(\mathcal{W}_i, \mathcal{Z}_J)$ we have 
\begin{equation}
\begin{split}
\mathscr{F}(\mathcal{W}_i, \mathcal{Z}_J) = \int \,&\frac{\mathrm{d}^{4k|4k}\mathcal{A} \; \mathrm{d}^{n+2}\sigma \; \mathrm{d}^{n+2}\xi}{\vol GL(2)} \, \frac{\prod_{J\in \mathsf{m}} \delta^{(4|4)}(\mathcal{Z}_J-\xi_J\mathcal{P}(\sigma_J, \{\cA\}))}{\xi_x \, \xi_y \, (\sigma_x-\sigma_y)^2 \; \prod_{a=1}^n \xi_a \, (\sigma_a-\sigma_{a+1})} \\
&\times \bigg\langle \frac{\partial}{\partial\mathcal{W}_x} \, \mathcal{I} \, \frac{\partial}{\partial\mathcal{W}_y} \bigg\rangle^{\!2} \, \prod_{i\in \bar{\mathsf{p}} } \exp\big( i \, \xi_i \, \mathcal{W}_i \cdot\mathcal{P}( \sigma_i, \{\mathcal{A}\} )\big) \ .
\end{split}
\end{equation}

{\bf 3.} The key observation of \cite{Spradlin:2009qr} is that this procedure has the advantage that there are now as many $\delta$-functions as moduli, and the integration over the $\mathcal{A}$ can be performed explicitly, with the net effect of localising the polynomial $\mathcal{P}(\sigma, \{\mathcal{A}\})$ onto 
\begin{equation}\label{Psol}
\mathcal{P}(\sigma) = \sum_{J\in \mathsf{m}} \frac{\mathcal{Z}_J}{\xi_J} \prod_{K \neq J} \frac{\sigma_K - \sigma}{\sigma_K - \sigma_J} \ .
\end{equation}
One is then left with
\begin{equation}
\begin{split}
\label{inter}
\mathscr{F}(\mathcal{W}_i, \mathcal{Z}_J) = \int \, &\frac{1}{\vol GL(2)} \, \frac{\mathrm{d}\sigma_x \, \mathrm{d}\xi_x \, \mathrm{d}\sigma_y \, \mathrm{d}\xi_y}{\xi_x \, \xi_y \, (\sigma_x-\sigma_y)^2} \, \prod_{a=1}^{n} \frac{\mathrm{d}\sigma_a \, \mathrm{d}\xi_a}{\xi_a \, (\sigma_a-\sigma_{a+1})} \\
& \times \bigg\langle \frac{\partial}{\partial\mathcal{W}_x} \, \mathcal{I} \, \frac{\partial}{\partial\mathcal{W}_y} \bigg\rangle^{\!2} \exp\bigg( i \sum_{i\in\bar{\mathsf{p}}, J\in\mathsf{m}} \!\! \mathcal{W}_i \cdot\mathcal{Z}_J \; \frac{\xi_i}{\xi_J} \prod_{K\neq J} \frac{\sigma_K - \sigma_i}{\sigma_K - \sigma_J}\bigg) \ .
\end{split}
\end{equation}

{\bf 4.} The change of variables
$(\xi_i, \xi_J )\to (t_i, t_J)$ with \cite{Spradlin:2009qr}
\begin{equation}\label{cov}
t_i \defeq \xi_i \prod_K (\sigma_K - \sigma_i) \, , \qquad \qquad t_J^{-1} \defeq \xi_J \prod_{K \neq J} (\sigma_K - \sigma_J) \, , 
\end{equation}
simplifies \eqref{inter} to
\begin{equation}\label{interbis}
\begin{split}
\mathscr{F}(\mathcal{W}_i, \mathcal{Z}_J) = \int\!&\frac{1}{\vol GL(2)} \, \frac{\mathrm{d}\sigma_x \, \mathrm{d}t_x \, \mathrm{d}\sigma_y \, \mathrm{d}t_y}{t_x \, t_y \, (\sigma_x-\sigma_y)^2} \, \prod_{a=1}^{n} \frac{\mathrm{d}\sigma_a \, \mathrm{d}t_a}{t_a \, (\sigma_a-\sigma_{a+1})} \\ 
&\times \bigg\langle \frac{\partial}{\partial\mathcal{W}_x} \, \mathcal{I} \, \frac{\partial}{\partial\mathcal{W}_y} \bigg\rangle^{\!2} \exp\bigg( i \sum_{i\in\bar{\mathsf{p}}, J\in\mathsf{m}} \!\! \mathcal{W}_i \cdot \mathcal{Z}_J \; \frac{t_i \, t_J}{\sigma_J - \sigma_i} \bigg) \, .
\end{split}
\end{equation} 
We now introduce spinor coordinates $\sigma_\alpha = t^{-1}(1, \sigma)$, with $(a\, b) \defeq \epsilon_{\alpha \beta} \sigma^{\alpha}_a \sigma^{\beta}_b$. Then  $\mathrm{d}^2 \sigma = t^{-3} \, \mathrm{d}t \, \mathrm{d}\sigma$, and 
\begin{equation}
\prod_{a=1}^n \frac{\mathrm{d}t_a \, \mathrm{d}\sigma_a}{t_a ( \sigma_a- \sigma_{a+1})} = \prod_{a=1}^n \frac{\mathrm{d}^2 \sigma_a}{(a \, a+1)} \, . 
\end{equation}
We then arrive at the very simple result
\begin{equation}
\begin{split}
\label{inter2}
\mathscr{F}(\mathcal{W}_i, \mathcal{Z}_J) &= \int\!\frac{1}{\vol GL(2)} \frac{\mathrm{d}^2\sigma_x \; \mathrm{d}^2\sigma_y}{(x \, y)^2} \, \prod_{a=1}^n \frac{\mathrm{d}^2 \sigma_a}{(a \, a+1)} \, \bigg\langle \frac{\partial}{\partial\mathcal{W}_x} \, \mathcal{I} \, \frac{\partial}{\partial\mathcal{W}_y} \bigg\rangle^{\!2} \! \exp\bigg(i \!\! \sum_{i\in\bar{\mathsf{p}}, J\in\mathsf{m}} \!\! \frac{\mathcal{W}_i \cdot \mathcal{Z}_J}{(i \, J)}\bigg) \, .
\end{split}
\end{equation}

{\bf 5.} It is now easy to go back to spinor variables by performing a Fourier transform. The result is%
\footnote{We recall that $\mathcal{Z} = (\lambda, \mu, \eta)$ and $\mathcal{W}=(\tilde{\mu}, \tilde{\lambda}, \tilde{\eta})$, with $\cZ\cdot \cW:=\lambda^\alpha \tilde{\mu}_\alpha+\mu_{\dot\alpha}\tilde{\lambda}^{\dot\alpha} + \tilde\eta_A\eta^A$.
}
\begin{align}
\label{FF_rational_SE}
\begin{split}
\mathscr{F}(\{\lambda,\tilde{\lambda}\}) = \langle x \, y \rangle^2 \int\!&\frac{1}{\vol GL(2)} \frac{\mathrm{d}^2\sigma_x \; \mathrm{d}^2\sigma_y}{(x \, y)^2} \prod_{a=1}^n \frac{\mathrm{d}^2 \sigma_a}{(a \, a+1)} \\
  & \times \prod_{i\in\bar{\mathsf{p}}} \delta^{(2)}(\lambda_i-\lambda(\sigma_i)) \prod_{J\in\mathsf{m}} \delta^{(2|4)}(\tilde{\lambda}_J-\tilde{\lambda}(\sigma_J), \eta_J-\eta(\sigma_J)) \,, 
\end{split}
\end{align}
 where we have defined the functions
\begin{equation}\label{scatfuncs}
\lambda(\sigma)  \, \defeq\, \sum_{J\in\mathsf{m}} \frac{\lambda_J}{(\sigma \, \sigma_J)} \,, \qquad \tilde{\lambda}(\sigma)  \, \defeq\,  \sum_{i\in\bar{\mathsf{p}}} \frac{\tilde{\lambda}_i}{(\sigma_i \, \sigma)} \,, \qquad \eta(\sigma) \, \defeq\,  \sum_{i\in\bar{\mathsf{p}}} \frac{\eta_i}{(\sigma_i \, \sigma)} \,.
\end{equation}
\eqref{FF_rational_SE} is nothing but the supersymmetric form of the scattering equation for form factors presented in \cite{He:2016dol}.  
By performing in reverse the same steps of this proof, one can of course derive the connected prescription for form factors \eqref{conFF} from the scattering equations.%
\footnote{We comment that the four-dimensional version of the amplitude scattering equations was noted in \cite{Spradlin:2009qr}.}

\section{A link representation for form factors}
\label{link-rep}

In \cite{Spradlin:2009qr}, Spradlin and Volovich presented an interesting formula for the $S$-matrix of $\cN\!=\!4$ SYM using the link variables introduced in \cite{ArkaniHamed:2009si}, and we now give the corresponding formula for the form factors of the stress tensor multiplet operator.

The link representation is obtained by introducing auxiliary variables
\begin{equation}
c_{iJ} \defeq \frac{1}{(i \, J)} \ ,
\end{equation}
where we note that the first and second index run over the sets $\bar{\mathsf{p}}$ and $\mathsf{m}$, respectively. This identification is achieved by introducing $1 = \int\!\mathrm{d}c_{iJ} \, \delta\left(c_{iJ} - 1 / (iJ)\right)$. Doing so, we can recast \eqref{FF_rational_SE} as
\begin{equation}
\label{linkrep}
\mathscr{F}(\{\lambda,\tilde{\lambda}\}) = \langle x \, y \rangle^2 \int \!\! \prod_{i\in\bar{\mathsf{p}},  J\in\mathsf{m}} \! \mathrm{d}c_{iJ} \; U(c_{iJ}) \, \prod_{i\in\bar{\mathsf{p}}} \delta^{(2)}(\lambda_i-c_{iJ}\lambda_J) \prod_{J\in\mathsf{m}} \delta^{(2|4)}(\tilde{\lambda}_J+c_{iJ}\tilde{\lambda}_i, \eta_J+c_{iJ}\eta_i) \,,
\end{equation}
where
\begin{equation}
\label{UUU}
U(c_{iJ}) \defeq \int \frac{1}{\vol GL(2)} \frac{\mathrm{d}^2\sigma_x \; \mathrm{d}^2\sigma_y}{(x \, y)^2} \, \prod_{a=1}^n \frac{\mathrm{d}^2 \sigma_a}{(a \, a+1)} \prod_{i\in\bar{\mathsf{p}}, J\in\mathsf{m}} \delta\Big(c_{iJ} - \frac{1}{(i \, J)}\Big) \,.
\end{equation}
There are several reasons why it is interesting to study the link representation form  \eqref{linkrep}. Firstly, it has the advantage of linearising momentum conservation in terms of the $c_{iJ}$ variables. Secondly, the quantity $U (c_{iJ})$ defined in \eqref{UUU} appears to be much more easily computed for any $k$ --  this is a considerable advantage in comparison to the scattering equations. Finally, it was shown in \cite{Spradlin:2009qr} that by using the (global) residue theorem, one can arrive at an alternative representation of the amplitudes which precisely matches BCFW diagrams, thus establishing a direct connection between the twistor-string representation of amplitudes and on-shell recursion relations. We will see that the same is also true for our representation of form factors, as we will explain in Section \ref{examples}. 

In performing explicit calculations, a natural way to fix the $GL(2)$ gauge freedom is to fix the four variables corresponding to the two auxiliary legs $x$ and $y$, 
\begin{equation}\label{gfix}
\sigma_x = (1,0) \,, \qquad \sigma_y = (0,1) \,, \qquad (x\,y) = 1 \,.
\end{equation}
We can then change variables from the spinors $\sigma_a^\alpha$ to the brackets $(x \, a)$ and $(y \, a)$ so that
\begin{equation}
\label{Ufix}
U(c_{iJ}) = \int \prod_{a=1}^n \frac{\mathrm{d} (x \, a) \; \mathrm{d} (y \, a)}{(a \, a+1)} \prod_{i\in\bar{\mathsf{p}}, J\in\mathsf{m}} \delta\Big(c_{iJ} - \frac{1}{(i \, J)}\Big)  \,.
\end{equation}
All the other  brackets can be  obtained from those used as integration variables using the  Schouten identity, 
\begin{equation}
(x \, y)(a \, b) = (x \, a) (y \, b) - (y \, a) (x \, b) \,.
\end{equation}
In \eqref{Ufix} we have $2n$ integration variables and $k(n+2-k)$ delta functions, which means that $U(c_{iJ})$ contains $(k-2)(n-k)$ delta functions after integration. In \eqref{linkrep}, four of the Grassmann-even delta functions  enforce momentum conservation, leaving  $2n$ delta functions and $k(n+2-k)$ variables $c_{iJ}$ to integrate over. This leaves $(k-2) (n-k)$ integration variables, which we denote by $\tau_k$.
Thus  \eqref{linkrep} can be written as
\begin{equation}\label{link_last_step}
\mathscr{F}(\{\lambda,\tilde{\lambda}\}) = J \, \langle x \, y \rangle^2 \, \delta^{(4)} \Big( q-\sum_{a=1}^n p_a\Big) 
\int \mathrm{d}^{(k-2)(n-k)}\tau \; U(c_{iJ}) \, \prod_{J\in\mathsf{m}} \delta^{(4)}(\eta_J+c_{iJ}\eta_i) \,,
\end{equation}
for some $c_{iJ}(\tau)$ linear in $\tau$, and an appropriate Jacobian $J$.

\section{Examples}
\label{examples}
In this section we work out explicitly the form of $U(c_{iJ})$ defined in \eqref{UUU} for various form factors. We will always use the gauge fixing \eqref{gfix}  so that $U(c_{iJ})$ is computed using  \eqref{Ufix}.

\subsection{The maximally non-MHV form factor}
The simplest form factor to discuss turns out to be  the maximally non-MHV form factor, where   $\bar{\mathsf{p}}=\{x,y\}$ and $\mathsf{m}=\{1,\ldots,n\}$. This corresponds to the case where for the component operator $\mathcal{L}_{\text{on-shell}}$ the on-shell state contains $n$ gluons all with negative helicity. In this case we find  
\begin{equation}
\label{callmi}
U^\textrm{max non-MHV} = \prod_{J=1}^n \frac{1}{c_{xy;J \, J+1}} \,,
\end{equation}
where we have defined
\begin{equation}
c_{ab; cd} \defeq c_{ac}\,c_{bd} - c_{ad}\,c_{bc} \,.
\end{equation}
 The expression for the $n$-point maximally non-MHV form factor is 
\begin{equation}
\begin{split}
\label{allms}
\mathscr{F}^\textrm{max non-MHV} =\ & \langle x \, y \rangle^2 \int \prod_{J=1}^n \mathrm{d}c_{xJ} \, \mathrm{d}c_{yJ} \; \delta^{(2)} \Big(\lambda_x - \sum_{J=1}^n c_{xJ} \lambda_J\Big) \; \delta^{(2)} \Big(\lambda_y - \sum_{J=1}^n c_{yJ} \lambda_J\Big) \\
&\times \prod_{J=1}^n \delta^{(2|4)}(\tilde{\lambda}_J + c_{xJ}\tilde{\lambda}_x + c_{yJ}\tilde{\lambda}_y, \eta_J + c_{xJ}\eta_x + c_{yJ}\eta_y) \times U^\textrm{max non-MHV} \ ,
\end{split}
\end{equation}
with $U^\textrm{max non-MHV} $ given in \eqref{callmi}.
We can first extract momentum conservation, 
\begin{equation}
\delta^{(2)} \Big(\lambda_x - \sum_{J=1}^n c_{xJ} \lambda_J\Big) \; \delta^{(2)} \Big(\lambda_y - \sum_{J=1}^n c_{yJ} \lambda_J\Big) = [x \, y]^2 \, \delta^{(4)} \Big( q-\sum_{J=1}^n p_J\Big) \ , 
\end{equation}
where we used $p_x + p_y = q$. It is then immediate to solve for the variables $c_{xJ}$ and $c_{yJ}$. The result is
\begin{equation}
c_{xJ} = \frac{[J \, y]}{[y \, x]} \ , \qquad \qquad c_{yJ} = \frac{[J \, x]}{[x \, y]}
\ . 
\end{equation}
The Jacobian from the delta functions for the $\tilde{\lambda}$ variables in \eqref{allms} contributes a factor of $[x \, y]^{-n}$, 
while 
a short calculation  shows that 
\begin{equation}
\label{cab}
c_{xy; J J+1} = \frac{[J \, J+1]}{[x \, y]} \ .
\end{equation}
Combining the Jacobian with the expression for \eqref{callmi} evaluated on \eqref{cab} leads to 
the known supersymmetric expression from \cite{Brandhuber:2011tv}, 
\begin{equation}
\label{qui} 
\mathscr{F}^\textrm{max non-MHV} = \frac{q^4}{[1 \, 2] \cdots [n \, 1]} \prod_{J=1}^n \delta^{(4)} \Big( \eta_J + \frac{[J \, y]}{[y \, x]} \eta_x + \frac{[J \, x]}{[x\, y]} \eta_y \Big)
\ . 
\end{equation}
Note that the fermionic delta function in \eqref{qui} imposes supermomentum conservation $\sum_{J=1}^n \lambda_J \eta_J = \lambda_x \eta_x + \lambda_y \eta_y$ (and we have dropped the obvious momentum conservation delta function). 
In particular the form factor of $\Tr (F_{\rm SD}^2)$ can be obtained by setting $\eta_x=\eta_y=0$ \cite{Brandhuber:2011tv}. For a state consisting only of gluons with negative helicity one obtains immediately the known result $q^4 / ([1\, 2] \cdots [n\, 1])$.

It is interesting to contrast the remarkable simplicity of this derivation from link variables  with the original calculation presented in \cite{Brandhuber:2011tv} of this component form factor, which required a more significant amount of work.

\subsection{The MHV form factors}
The next simplest case is that of an MHV form factor. In this case the two sets are  $\bar{\mathsf{p}}=\{1,\ldots, \widehat{J_1}, \ldots, \widehat{J_2}, \ldots,n,x,y\}$ and $\mathsf{m}=\{J_1, J_2\}$, where hatted entries are to be omitted from the set. This corresponds to the case where, for the component operator $\mathcal{L}_{\text{on-shell}}$ and a purely gluonic on-shell state,  there are exactly two gluons with negative and $n-2$ with positive helicity.

The $U$ function in \eqref{Ufix} is given by 
\begin{equation}
U^\textrm{MHV}=\frac{1}{(c_{xy;J_1 J_2})^2 \, c_{J_1-1\,J_2} \, c_{J_1+1\,J_2} \, c_{J_1\,J_2-1} \, c_{J_1\,J_2+1}} \; \prod_{a \neq J_1-1,J_1,J_2-1,J_2} \frac{1}{c_{a \, a+1;J_1 J_2}} \ .
\end{equation}
Performing the integration over the link variables is again straightforward,  and one arrives at the MHV super form factor of the chiral
part of the stress tensor multiplet
\begin{equation}
\label{qui2} 
\mathscr{F}^\textrm{MHV} = \frac{1}{\langle 1 \, 2\rangle \cdots \langle n \, 1 \rangle} \; \delta^{(8)} \Big( \sum_{a=1}^n  \lambda_a \eta_a+ \lambda_x \eta_x+  \lambda_y \eta_y\Big) \, , 
\end{equation}
which agrees with the known result \cite{Brandhuber:2011tv} if we identify $\gamma_+ = \lambda_x \eta_x+ \lambda_y\eta_y $.

\subsection{The $1^-2^-3^-4^+$ form factor and connection to BCFW diagrams}\label{BCFW-sec}
In this case we have $I=\{1,2,3\}$ and $i=\{4,x,y\}$,  and \eqref{Ufix} reads
\begin{equation}
\label{Ufix6point}
U^{1^- 2^- 3^-4^+} = \int \prod_{a=1}^4 \frac{\mathrm{d} (x \, a) \; \mathrm{d} (y \, a)}{(a \, a+1)} \prod_{J=1}^{3} \delta\Big(c_{xJ} - \frac{1}{(x\, J)}\Big) \, \delta \Big(c_{yJ} - \frac{1}{(y\, J)}\Big)\, \delta \Big(c_{4J} - \frac{1}{(4\, J)}\Big) \,.
\end{equation}
With nine delta functions and eight integrations, there is one delta function remaining after all integrations are carried out. The integrations over $(x\, J)$ and $(y\, J)$ are straightforward, and one can then choose to solve the two delta functions involving $(4\, 1)$ and $(4\, 2)$, producing a Jacobian, and insert this solution into the remaining delta function for $(4\, 3)$. Collecting all terms from this process, one finds that  that
\begin{equation}\label{Ummmp}
U^{1^- 2^- 3^-4^+} = \frac{c_{x2} \, c_{y2}}{c_{42} \, c_{xy; 21} \, c_{xy; 23}} \; \delta(S_{123;4xy}) \,, 
\end{equation}
where, following the notation introduced in \cite{Spradlin:2009qr}, we define
\begin{equation}
S_{ijk;lmn} \defeq c_{mi} \, c_{mj} \, c_{lk} \, c_{nk} \, c_{ln;ij} - c_{ni} \, c_{nj} \, c_{lk} \, c_{mk} \, c_{lm;ij} - c_{li} \, c_{lj} \, c_{mk} \, c_{nk} \, c_{mn;ij}
\ . 
\end{equation}
We comment that in this case $S_{ijk,LMN} = (\prod_{iJ}C_{iJ})\, {\rm det}([1/C_{iJ}])$ where $[1/C_{iJ}]$ is the matrix with elements $1/C_{iJ}$.
As in \eqref{link_last_step}, the form factor can be obtained by integrating out the remaining delta function. However, there is a more efficient way to derive the final result which avoids solving the constraint of $\delta(S_{123;4xy})$ altogether.

In general, the complex delta function has the property 
\begin{equation}
\int\!\mathrm{d} z_1 \ldots \mathrm{d} z_m \ g(z) \; \delta(f_1(z)) \cdots \delta(f_m(z)) = \! \sum_{z_0 \in f^{-1}(0)}\!\!\!\left. {\Res} \, \omega\right|_{z_0} \,,
\end{equation}
where 
\begin{equation}
\omega \defeq \frac{g(z) \, \mathrm{d} z_1 \wedge \ldots \wedge \mathrm{d} z_m}{f_1(z) \ldots f_m(z)} \,.
\end{equation}
In our case, this means that the integral in \eqref{link_last_step} can be written as a sum of residues of
\begin{equation}
\omega_U = \frac{c_{x2} \, c_{y2}}{c_{42} \, c_{xy; 21} \, c_{xy; 23}} \frac{\mathrm{d}\tau}{S_{123;4xy}}\, , 
\end{equation}
evaluated on the zeros of the quartic polynomial $S_{123;4xy}(\tau)$. However, since $\omega_U$ can be straightforwardly extended to a meromorphic form on $\mathbb{CP}^1$, we can use the global residue theorem to compute the result in terms of the other poles of $\omega_U$, which correspond to the simple zeros of $c_{42}(\tau)$, $c_{xy; 21}(\tau)$ and $c_{xy; 23}(\tau)$. Focusing on gluon scattering, the corresponding  residues are
\begin{equation}\label{bcfw-parts}
\begin{split}
\mathscr{F}_{42} &= - \frac{\langle 1 \, 3 \rangle \, q^4}{s_{134} \, \langle 1 \, 4 \rangle \, \langle 3 \, 4 \rangle \, \langle 3|q|2] \, \langle 1|q|2]} \,, \\
\mathscr{F}_{xy; 21} &= - \frac{\langle 3|q|4]^3}{s_{124} \, [1 \, 2] \, [1 \, 4] \, \langle 3|q|2]} \,, \\
\mathscr{F}_{xy; 23} &= - \frac{\langle 1|q|4]^3}{s_{324} \, [3 \, 2] \, [3 \, 4] \, \langle 1|q|2]} \,,
\end{split}
\end{equation}
and the complete result is obtained by adding the three terms, 
\begin{equation}\label{F1234_sum}
\mathscr{F}^{1^- 2^- 3^-4^+} = \mathscr{F}_{42} + \mathscr{F}_{xy; 21} + \mathscr{F}_{xy; 23} \,.
\end{equation}
It is notable that each term in \eqref{bcfw-parts} depends on $p_x$ and $p_y$ only through the combination $p_x+p_y = q$. Moreover, each term is a rational function of external kinematics. Interestingly, these two properties do not hold for the four terms arising from the solutions of the scattering equation $S_{123;4xy} = 0$, and are only recovered in the sum over the four solutions.

Perhaps more remarkably, each term in \eqref{bcfw-parts} corresponds to a BCFW diagram for a $[1 \ 2\rangle$ shift, analogously to the amplitude case, as discussed in \cite{Spradlin:2009qr}. Specifically, we have found that the sum in \eqref{F1234_sum} corresponds, term by term, to the sum
\begin{equation}
\raisebox{-.45\height}{\includegraphics{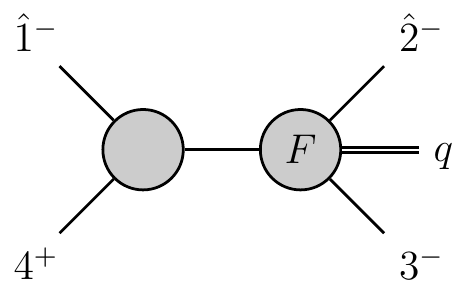}}
\; + \;
\raisebox{-.45\height}{\includegraphics{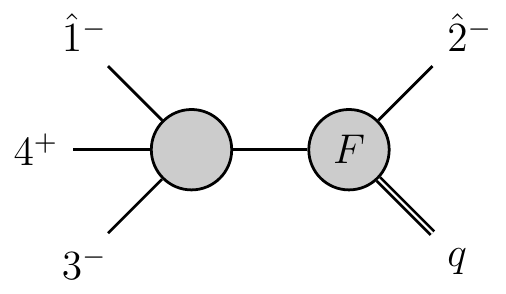}}
+
\raisebox{-.45\height}{\includegraphics{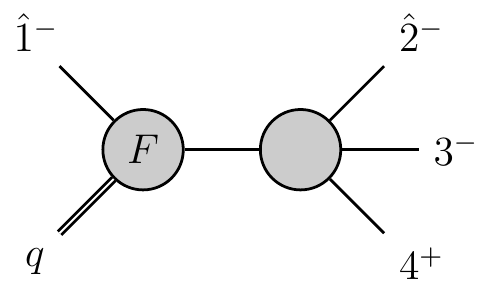}}
\end{equation}
given by the BCFW expansion of the form factor.

\section{Connections to the Grassmannian}
\label{grasss}
In \cite{Frassek:2015rka} it was conjectured that the $n$-point N$^{k-2}$MHV chiral stress-tensor super form factor in $\mathcal{N}\!=\!4$ SYM can be obtained as an integral over the Grassmannian\footnote{$G(k,n)$ is defined to be the set of $k$-dimensional linear subspaces of $\mathbb{C}^n$.} $G(k,n+2)$ of the form
\begin{equation}
\begin{split}
\langle n\!+\!1 \, n\!+\!2 \rangle^2 \int \frac{\mathrm{d}^{k(n+2)}C \ \mathrm{d}^{2k} \rho}{\vol GL(k)}\ &\sum_{\textrm{ins}} \frac{\Omega_{n,k}(C)}{(1 \cdots k ) \cdots (n+2 \cdots k-1)}
\\
&\times \delta^{2k}(C \cdot \tilde\lambda)\, \delta^{2(n+2)}(\rho \odot C -  \lambda )\, \delta^{4k}(C \cdot \eta) \, ,
\label{wilhelm-conj}
\end{split}
\end{equation}
where in this section the auxiliary particles $x$ and $y$ are denoted as $n+1$ and $n+2$, respectively. 
$C_{I a}$ is a $k\times (n+2)$ matrix, describing a $k$-plane in $\mathbb{C}^{n+2}$, $\rho_J^\alpha$ is a $k\times 2$ matrix, $\cdot$ indicates contraction of the $a$ indices, $\odot$ denotes contraction of the $J$ indices%
\footnote{For clarity, $a=1,\ldots, n+2$ and $J \in \mathsf{m}$. In this section we will also set $\mathsf{m} = \{1, \ldots, k\}$ and $\mathsf{p} = \{k+1, \ldots, n\}$ for convenience.}, $(a \cdots b)$ is a $k\times k$ minor of $C$, and the sum runs over allowed insertions of $\{ n+1, n+2 \}$ into $\{1, \dots n \}$.
The integrand $\Omega_{n,k}(C)$ appearing in \eqref{wilhelm-conj} is defined to be
\begin{equation}
\Omega_{n,k}(C) = \frac{Y}{1-Y} \, , \quad \qquad Y = \frac{(n+2-k \cdots n \ n+1)(n+2 \ 1 \cdots k-1)}{(n+2-k \cdots n\ n +2 )(n+1\ 1 \cdots k-1)} \, .
\end{equation}
We are interested in connecting this conjecture with the result \eqref{FF_rational_SE} expressing the form factor as a sum over solutions to the rational scattering equations. An obvious approach is suggested by comparison to \cite{ArkaniHamed:2009dg}, in which Grassmannian amplitude formulae are mapped to CHY-type formulae courtesy of the Veronese map. The Veronese map is an embedding of $G(2,n+2)$ in $G(k,n+2)$ defined by taking
\begin{equation}
C_{J a} = \xi_a \, \sigma_a^{J-1} \, .
\end{equation}
We can perform a partial integration of \eqref{wilhelm-conj}, reducing it to an integral over $G(2,n+2)$ in this embedding:\footnote{Naively one might worry that there are four fewer integration variables than $\delta$-functions. However the leftover constraints combine to form the $\delta$-function of momentum conservation in the final answer, as required.}
\begin{equation}
\begin{split}
\langle n\!+\!1 \, n\!+\!2 \rangle^2 \int &\, \frac{\rd^{n+2}\sigma \, \rd^{n+2}\xi \, \rd^{2k} \rho}{\vol GL(2)} \sum_{\textrm{ins}}\frac{\Omega^V_{n,k}(\sigma_a, \xi_a)}{\prod_{a=1}^{n+2}\xi_a (\sigma_a - \sigma_{a+1})}
\\
&\times \prod_{a=1}^{n+2} \delta^{(2)}\Big(\lambda_a - \xi_a \sum_{J\in \mathsf{m}}\rho_{J}\sigma_a^{J-1}\Big) \prod_{J\in \mathsf{m}} \delta^{(2|4)}\Big(\sum_{a=1}^{n+2} \xi_a \sigma_a^{J-1} \{\tilde \lambda_a|\eta_a\}\Big)
\, .
\end{split}
\end{equation}
The $\delta$-functions enforce the polynomial scattering equations in the language of \cite{He:2016vfi}. For practical purposes it it more convenient to gauge fix the $GL(k)$ symmetry before applying the Veronese map, enforcing the rational scattering equations of \cite{Geyer:2014fka}. We also apply the change of variables \eqref{cov} to yield
\begin{equation}
\begin{split}
\label{our-conj}
\langle n\!+\!1 \, n\!+\!2 \rangle^2 \int & \frac{\rd^{2(n+2)}\sigma}{\vol GL(2)} \sum_{\textrm{ins}}\frac{\Omega^V_{n,k}(\sigma_a,t_a)}{(\sigma_1 \sigma_2)\cdots (\sigma_{n+2}\sigma_1)}
\\
&\times \prod_{i \in \bar{\mathsf{p}}} \delta^{(2)} (\lambda_i - \lambda(\sigma_i)) \prod_{J \in \mathsf{m}} \delta^{(2|4)} (\tilde \lambda_J - \tilde \lambda (\sigma_J), \eta_J - \eta(\sigma_J)) \, ,
\end{split}
\end{equation}
where the functions defining the scattering equations are given by \eqref{scatfuncs}.

Under the Veronese map, $Y$ then becomes
\begin{equation}\label{Veroneasy}
Y^V(\sigma_a, \xi_a) = \prod_{j=n+2-k}^n \frac{\sigma_j - \sigma_{n+1}}{\sigma_j - \sigma_{n+2}} \; \prod_{i=1}^{k-1}\frac{ \sigma_{n+2} - \sigma_i}{ \sigma_{n+1} - \sigma_i} \ ,
\end{equation}
after using the Vandermonde determinant formula. Note immediately that this is independent of the $ \xi_a$, thus the transition to the rational scattering equation version is simply the identity map.

In relation to \eqref{wilhelm-conj}, with $Y_k = Y^V(\sigma_a, \xi_a)$ and $C_n = (\sigma_1-\sigma_2)(\sigma_1-\sigma_2)\cdots(\sigma_n-\sigma_1)$ it may be of interest to note the following recursive expression
\begin{equation}\label{Yrecursion}
\frac{1}{C_{n+2}} \ \frac{Y_k}{1-Y_k} = \frac{1}{{(\sigma_{n+1}-\sigma_{n+2})^2 C_n}} \ \frac{N_k}{D_k} \ , 
\end{equation}
with
\begin{equation}
N_k = (\sigma_n-\sigma_1)\prod_{i=1}^{k-2} (\sigma_{n-i}-\sigma_{n+1})(\sigma_{n+2}-\sigma_{n+3+i}) \ ,
\end{equation}
and
\begin{equation}
D_k = N_k + (\sigma_n-\sigma_{n+2})(\sigma_{n+1}- \sigma_1) D_{k-1}(\hat \sigma_n,\hat \sigma_1) \ ,
\end{equation}
with $D_{k-1}(\hat \sigma_n,\hat \sigma_1)$ meaning $D_{k-1}$ with variables $\sigma_1, \sigma_2,\dots,\sigma_{n+2}$ but omitting $\sigma_1$ and $\sigma_n$.
We have checked \eqref{Yrecursion} algebraically with {\tt Mathematica} up to $k=10$ and numerically for various higher values.

In terms of the homogeneous coordinates, \eqref{Veroneasy} is
\begin{equation}\label{from-grassmannian}
Y^V(\sigma_a) = \prod_{j=n+2-k}^n \frac{(j \, n+1)}{(j \, n+2)} \; \prod_{i=1}^{k-1} \frac{(n+2 \, i)}{(n+1 \, i)} \ . 
\end{equation}
The authors of \cite{He:2016dol} conjectured a simpler formula for the chiral stress tensor super form factor, namely \eqref{FF_rational_SE}. 
In the cases $k = 2$ and $k = n$ a short calculation shows agreement with the formula \eqref{from-grassmannian} obtained from the Grassmannian. Indeed, these cases correspond to the MHV and maximally non-MHV form factor, where the sum in \eqref{wilhelm-conj} consists of a single term. More generally, one must sum over terms arising from several top-cell forms constructed via on-shell diagrams. These correspond to particular cyclic shifts of the insertion point of the additional legs representing the form factor.

The first non-trivial case in which we wish to show agreement between \eqref{our-conj} and \eqref{FF_rational_SE} is $n=4$, $k=3$, which corresponds to the helicity assignment $1^-2^-3^-4^+$ in our chosen convention. For this case it was shown in \cite{Frassek:2015rka} that the appropriate insertions are $\{1,2,3,4,5,6\}$ and $\{1,2,5,6,3,4\}$. A little algebra suffices to prove that 
\begin{align}
\begin{split}
\label{Y1}
&\frac{Y_1}{1-Y_1} \frac{1}{(1 \ 2 )(2 \ 3)(3 \ 4)(4 \ 5)(5 \ 6)(6 \ 1)} + \frac{Y_2}{1-Y_2}\frac{1}{(1 \ 2)(2 \ 5)(5 \ 6)(6 \ 3)(3 \ 4)(4 \ 1)}\\
&=\frac{1}{(1 \ 2)(2 \ 3)(3 \ 4)(4 \ 1)(5 \ 6)^2 } \ ,
\end{split}
\end{align}
where
\begin{equation}
Y_1 = \frac{(3 \ 5)(4 \ 5)(6 \ 1)(6 \ 2)}{(3 \ 6)(4 \ 6)(5 \ 1)(5 \ 2)}\, , \qquad \qquad Y_2 = \frac{(1 \ 5)(2 \ 5)(6 \ 3)(6 \ 4)}{(1 \ 6)(2 \ 6)(5 \ 3)(5 \ 4)}
\ .
\end{equation}
Note that in \eqref{Y1} we have obtained the expected integrand, where the auxiliary particles associated to the form factor now only appear in the factor $(5 \ 6)^2$.

The next non-trivial case is $n\!=\!5$, $k\!=\!3$. In this case, we have checked numerically that no combination of insertions reproduces the formula \eqref{FF_rational_SE}. This is not so surprising, since in this case different residues are required from each top-cell diagram, whereas the Veronese map treats terms democratically. It would be interesting to determine whether there is an improved choice of top-cells compatible with a Veronese reduction. We leave this question to future work.

\section{Form factors from ambitwistor strings}
\label{wild}

The result \eqref{FF_rational_SE} bears a close resemblance to the formula\footnote{Our superamplitudes have $\eta^0$ for positive helicity and $\eta^4$ for negative helicity gluons, which is the opposite of the convention employed in \cite{Geyer:2014fka}.}
\begin{equation}\label{Ampl_rational_SE}
\begin{aligned}
\mathscr{A}(\{\lambda,\tilde{\lambda}\}) = \int &\frac{1}{\vol GL(2)} \prod_{a=1}^n \frac{\mathrm{d}^2 \sigma_a}{(a \, a+1)} \\
  & \times \prod_{i\in\bar{\mathsf{p}}} \delta^{(2)}(\lambda_i-\lambda(\sigma_i)) \prod_{J\in\mathsf{m}} \delta^{(2|4)}(\tilde{\lambda}_J-\tilde{\lambda}(\sigma_J), \eta_J-\eta(\sigma_J)) \,,
\end{aligned}
\end{equation}
first derived in \cite{Geyer:2014fka} from an ambitwistor-string model, describing the tree-level $n$-particle scattering in four-dimensional $\mathcal{N}\!=\!4$ SYM.

In this construction the Parke-Taylor denominator of the measure emerges from a current algebra on the worldsheet, similarly to the standard heterotic string construction. Each vertex operator is dressed with a current $J^a$ built from $N$ free complex fermions $\psi^i$ and $SU(N)$ generators $T^a$. More explicitly, we define
\begin{equation}
J^a (\sigma) = \frac{i}{2} T_{ij}^a : \psi^i(\sigma) \bar \psi^j(\sigma) : \, ,
\end{equation}
where $i,j$ are fundamental representation indices and $a$ is an adjoint representation index. Recall that the only non-vanishing Wick contraction between complex fermions takes the form
\begin{equation}
\langle \psi^i (\sigma_1) \bar\psi^j (\sigma_2) \rangle = \frac{\delta^{ij}}{\sigma_1 - \sigma_2} \, ,
\end{equation}
so we may immediately evaluate the correlator of $n$ currents to be
\begin{equation}
\langle J^{a_1} \cdots J^{a_n} \rangle = \frac{\Tr\left(T^{a_1} \cdots T^{a_n} \right)}{(\sigma_1 - \sigma_2) \cdots (\sigma_n - \sigma_1) } + {\rm perms} + \cdots\, ,
\end{equation}
where we have ignored multiple trace terms. Keeping only the first term corresponds to computing a certain colour-ordered amplitude. 

We may construct the measure of formula \eqref{FF_rational_SE} from ambitwistor strings in a similar way, at least up to an overall factor. We must include two additional vertex operators, corresponding to the punctures $\sigma_{n+1}$ and $\sigma_{n+2}$ on the Riemann sphere. These are dressed with additional currents defined as above. However, in order to obtain the chiral stress tensor super form factor, we now do not require the single trace term. Rather we extract from Wick's theorem the double trace term displayed below, 
\begin{equation}
\langle J^{a_1} \cdots J^{a_{n+2}} \rangle = \dots +\frac{\Tr\left(T^{a_1} \cdots T^{a_n} \right)}{(\sigma_1 - \sigma_2) \cdots (\sigma_n - \sigma_1) \, (\sigma_{n+1}-\sigma_{n+2})^2} \cdot \Tr \left(T^{a_{n+1}}T^{a_{n+2}}\right)
+ \cdots \, ,
\end{equation}
providing the appropriate denominator and colour factor for the on-shell state.
It would be very interesting to have a complete derivation of \eqref{FF_rational_SE} from ambitwistor strings, also explaining the $\lan x\, y\ran^2$ prefactor. 

Of course, the current algebra in the four-dimensional ambitwistor string construction is identical to that in the ten-dimensional formula of \cite{Mason:2013sva} which reproduces standard CHY formulae. We might thus recast the formula \eqref{FF_rational_SE} as a sum over solutions to the standard scattering equations \cite{Cachazo:2013hca}. To do this would require an appropriate prescription for the polarisation vector associated with the off-shell insertion. 

Given that form factors emerge so naturally from an ambitwistor string construction, it is tempting to speculate that appropriate current algebra modifications might allow the construction of still more general objects, namely correlation functions. 

An obvious generalisation of the approach followed for form factors would be to include additional auxiliary particles to represent further operator insertions. The simplest example would be that of a two-point correlator of $\cO =\Tr F_{\rm SD}^2$ with the vacuum as the external state. In order to contract the two operators, we choose the two pairs of auxiliary particles to have opposite helicity, $(x^+, y^+)$ and $(u^-, v^-)$. The corresponding quantity is
\begin{equation}
\label{corr}
\begin{split}
 \lan x\, y\ran^2 \, [u\, v]^2
\int &\frac{1}{\vol GL(2)} \frac{\mathrm{d}^2\sigma_x \; \mathrm{d}^2\sigma_y}{(x \, y)^2} \frac{\mathrm{d}^2\sigma_u \; \mathrm{d}^2\sigma_v}{(u \, v)^2}
 \prod_{i=x,y} \delta^{(2)}(\lambda_i-\lambda(\sigma_i)) \prod_{J=u,v} \delta^{(2)}(\tilde{\lambda}_J-\tilde{\lambda}(\sigma_J)) \,, 
\end{split}
\end{equation}
with 
\begin{equation}
\lambda(\sigma) \, \defeq\, \sum_{J=u,v} \frac{\lambda_J}{(\sigma \, \sigma_J)} \,, \qquad \tilde{\lambda}(\sigma)  \, \defeq\, \sum_{i=x,y} \frac{\tilde{\lambda}_i}{(\sigma_i \, \sigma)} \, .
\end{equation}
An explicit calculation shows that \eqref{corr} is equal to
\begin{equation}
q^4 \, \delta^{(4)} (p_x + p_y +p_u + p_v) 
\, . 
\end{equation}
with $q=p_x + p_y$. This is not quite the result one expects to find for $\langle \cO (q) \bar\cO (q') \rangle$ from Equation (26) of \cite{Gubser:1998bc}, namely $\langle \cO (q) \bar\cO (q') \rangle \sim \delta^{(4)} ( q + q') \, q^4 \log ( q^2) \, + \, {\rm analytic \ terms}$. Note in particular that the $\log q^2$ term is absent. In order to be able to derive such terms one may need to understand scattering equations for off-shell quantities at loop level, along the lines of \cite{Geyer:2015bja,Geyer:2015jch,Geyer:2016wjx}.
It would also be very interesting to find concrete vertex operators in the ambitwistor string construction of \eqref{FF_rational_SE} that correspond to the operator insertions on the field theory side, possibly making contact with the the recent ideas in \cite{Koster:2016ebi,Chicherin:2016soh,Koster:2016loo, Chicherin:2016qsf}.
We hope to report on some of these ideas in a future publication.

\section*{Acknowledgements}
 
We would like to thank Paul Heslop and Brenda Penante for very interesting discussions.
This work was supported by the Science and Technology Facilities Council (STFC) Consolidated Grant ST/L000415/1
\textit{``String theory, gauge theory \& duality"}. The work of EH was supported by an STFC quota studentship.

\bibliographystyle{utphys}
\bibliography{twistor}
\end{document}